      [1999/12/01 v1.4c Il Nuovo Cimento]
\documentclass{cimento}

\newcommand{\lsim}{\ \raise -2.truept\hbox{\rlap{\hbox{$\sim$}}\raise5.truept
        \hbox{$<$}\ }}
\newcommand{\gsim}{\ \raise -2.truept\hbox{\rlap{\hbox{$\sim$}}\raise5.truept
        \hbox{$>$}\ }}

\newcommand{\grb}{GRB~020813}

\usepackage{graphicx}  % got figures? uncomment this
\title{The dust depletion and extinction of the GRB 020813 afterglow}
\author{S.~Savaglio\from{ins:JHU} \atque
S.~M.~Fall\from{ins:STScI}}
\instlist{\inst{ins:JHU} Johns Hopkins University - Baltimore, USA
  \inst{ins:STScI} Space Telescope Science Institute - Baltimore, USA}
\begin{document}
\maketitle

\begin{abstract}
The Keck optical spectrum of the GRB 020813 afterglow is the best ever
obtained for GRBs. Its large spectral range and very high S/N ratio
allowed for the first time the detection of a vast variety of
absorption lines, associated with the circumburst medium or
interstellar medium of the host.  The remarkable similarity of the
relative abundances of 8 elements with the dust depletion pattern seen
in the Galactic ISM suggests the presence of dust. The derived visual
dust extinction $A_V=0.40\pm0.06$ contradicts the featureless UV
spectrum of the afterglow, very well described by a unreddened power
law.  The forthcoming Swift era will open exciting opportunities to
explain similar phenomena in other GRB afterglows.
\end{abstract}

\section{Introduction}

The optical afterglow of GRB 020813 ($z=1.25$) was observed a few
hours after the burst with Keck at a reasonable spectral
resolution\cite{ref:barth} when the burst was still relatively
bright. The unprecedented\footnote{Much higher resolution spectra have been
obtained for GRB~020813 and GRB~021004\cite{ref:fiore}, however
the information available is somehow limited by the low S/N
in large parts of the spectrum.}
good quality of rest-frame UV spectrum, and the large wavelength range
covered, allowed the detection of a tremendous number of absorption
lines, including very weak transitions never detected before in GRB
spectra.

The GRB rest-frame UV spectrum shows another notable feature: the
continuum emission in the interval $7.4\times10^{14}-2.1\times10^{15}$ Hz
is very well represented by a perfect power law of the form $f_\nu
\propto
\nu^{-0.918\pm0.001}$\cite{ref:savaglio}. Such a 
nearly perfect power law is not only a very rare event, bust it also
strongly suggests (by the standard view of extinction laws) no
reddening, or equivalently, little dust along the sight-line.

Here we discuss the dust properties in the interstellar or circumburst
medium in front of GRB~020813 by using two different and completely
independent means: $i)$ studying the relative abundances of several
elements with very different refractory properties (i.e.\ how easily
they are locked in dust grains); $ii)$ trying to reconcile the
observed straight UV continuum with the presence of several extinction
laws.

\section{Metal abundances and visual extinction}

We derived column densities of FeII, SiII, SiII$^*$, ZnII, CrII, NiII,
MgI, CI, CaII, CIV, MnII, MgII, AlII, AlIII, and TiII in the
GRB~020813 afterglow, using independently line
fitting, curve of growth and apparent optical depth analysis
\cite{ref:savaglio}. High column densities of low ionization
species (e.g. FeII), relative to higher-ionization species (CIV and
AlIII), is a good indication that the gas is nearly neutral (the
column density of hydrogen is mostly in HI form) and the column
density of each element can be approximated by the column density of
the ion with ionization potential below 13.6 eV (e.g. Fe $\sim$
FeII). As the Lyman series lines are not in the observed spectral
range, we cannot estimate the HI column density. However, we can
derive relative metal abundances, compare them with solar values and
with relative abundances expected if the various elements are partly
locked in the dust grains as in the Milky Way ISM (dust depletion
pattern). Figure~1 shows with no ambiguity that relative abundances
can be explained if the gas is polluted by dust.

\begin{figure}
\centerline{\includegraphics{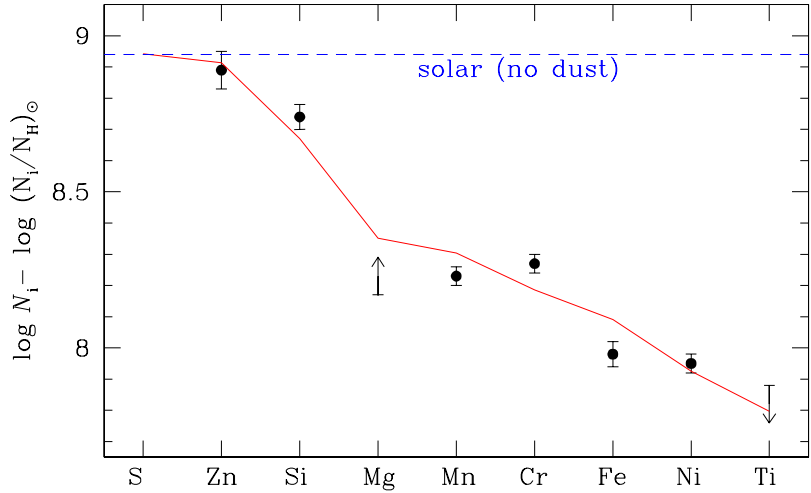}}
\caption{Heavy element column densities in \grb. The dashed line shows
solar relative abundances and no dust. The solid line shows the
expected column densities in the case that elements are partly locked
in dust grains, as in the gas of the warm disk+halo of the Milky
Way. Other depletion patterns in the MW have been considered, with
slightly worse fit (higher $\chi^2$).}
\vspace{-.3cm}
\end{figure}

The dust extinction of the GRB afterglow as a function of wavelength
depends basically on the dust column density and the grain-size
distribution.  While there is no obvious way we can determine the
latter, the former can be derived by assuming that the element pattern
observed in Figure~1 is a dust depletion pattern, with no need to know
the metal abundance relative to hydrogen.  The main assumption is that
the total (dust+gas) element relative abundance in the GRB ISM is as
in the solar neighborhood, and what is missing in the gas form is locked
into dust grains.

To estimate the visual extinction for GRB~020813, we also assume that
the rate of visual extinction per unit column of dust is as in the
solar neighborhood\cite{ref:bohlin}, or half a magnitude for a total
column density of Fe (dust and gas) $N_{\rm FeII}=10^{16.5}$
cm$^{-2}$.  From Figure~1, we estimated a total Fe column density
$N_{\rm FeII}=10^{16.46\pm0.08}$, and translated into a visual
extinction $A_V=0.40\pm0.06$.

\begin{figure}
\includegraphics{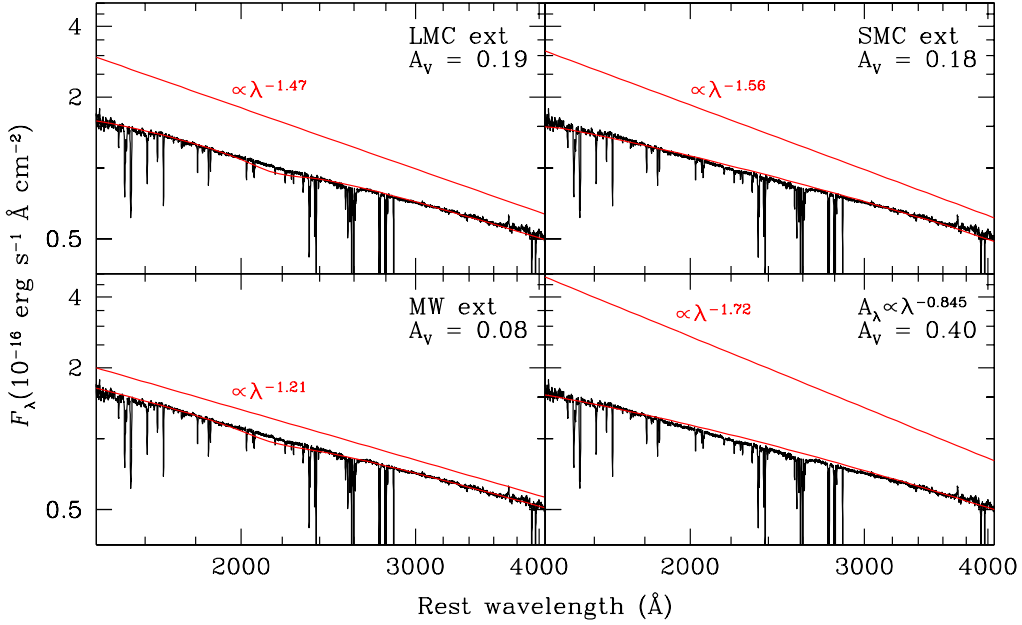}
\caption{GRB~020813 spectrum before and after a dust extinction screen 
is applied to an intrinsic GRB emission. The MW, LMC, SMC and
power-law extinctions are used. For the first three, $A_V$ is the
maximum value allowed by the observed spectrum. For the power-law
extinction, $A_V=0.4$ is assumed, while the power index $\gamma=0.845$
is the maximum value allowed by the observed GRB spectrum.}
\vspace{-.3cm}
\end{figure}

\section{The dust extinction law}

A possible extinction law for the ISM of the GRB has to be shallow
enough and/or a weak function of the wavelength in order to make the
GRB UV emission consistent with the observed undistorted power law. To
constrain the extinction law, we only assumed that the intrinsic GRB
emission is itself described by a (unconstrained) power law. We used
the MW, LMC, SMC extinctions, and a power-law extinction of the form
$\propto \lambda^{-\gamma}$.  Starting from all possible intrinsic GRB
emissions, we constrained $A_V$ for all four different
extinctions. Figure~2 shows for MW, LMC and SMC extinctions, the
maximum allowed $A_V$ still compatible with the observed GRB spectrum
(2 $\sigma$), together with the associated intrinsic emission. For the
MW and LMC extinctions, $A_V<0.08$ and $A_V<0.18$, respectively,
mainly due to the absence in the observed GRB spectrum of the 2200
\AA\ feature. For the power-law extinction, if $A_V=0.4$, then only
extinctions with $\gamma<0.85$ make the GRB spectrum consistent with
the observed one. This last result is summarized in Figure~3, where
the allowed regions of $A_V$ (from the depletion pattern analysis) and
$\gamma$ (from the extinction law analysis) are displayed together.

\begin{figure}
\centerline{\includegraphics{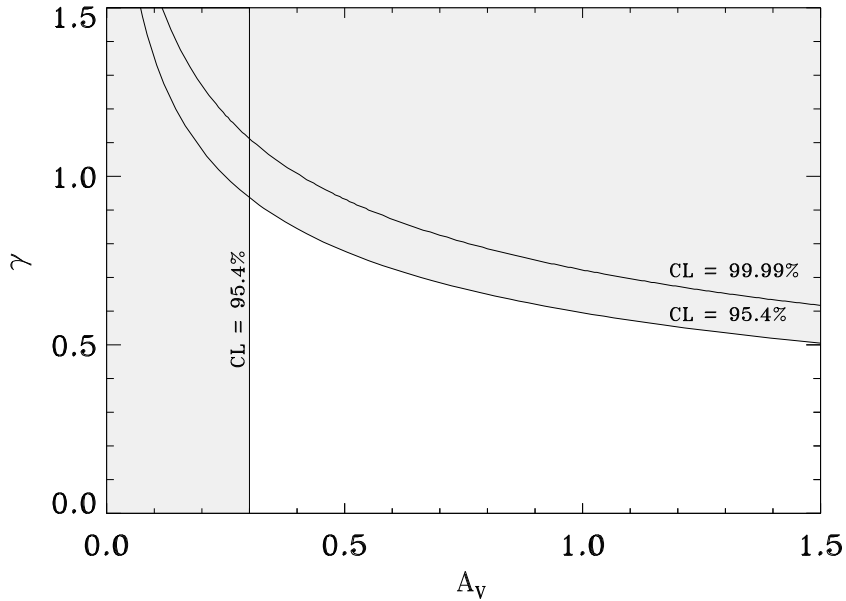}}
\caption{The shaded regions indicate the values of visual
extinction $A_V$ and of slope of the power-law extinction $\gamma$ not
allowed by the analysis of the dust properties in the GRB~020813
afterglow.}
\vspace{-.2cm}
\end{figure}

\section{Summary}

The relative metal abundances of 8 elements are derived with
unprecedented accuracy for a GRB afterglow.  These are remarkably
similar to the metal pattern observed in Galactic ISM clouds, polluted
by dust. If dust is indeed responsible of the metal pattern shown in
Figure~1, then the visual extinction is $A_V\simeq0.4$. Such a
relatively large extinction is compatible with the observed spectrum
(well represented by a power law of the form $f_\nu\propto
\nu^{-0.92}$) only assuming a shallow extinction law.  If we use the
known extinction laws (MW, LMC and SMC) the extincted spectrum is
still compatible with the observed spectrum if $A_V<0.2$. The observed
GRB spectral shape and the $A_V>0.3$ (95\% CL) derived from the
depletion pattern, can only be explained for a flatter extinction.
For instance, if a power law is assumed and $A_V=0.4$, than the slope
has to be shallower than $\gamma\simeq 0.85$.

GRB UV spectra will show their overwhelming potentiality for the
understanding of the cosmic chemical enrichment and physical state of
the high-$z$ interstellar medium when the Swift satellite will be
fully operational.  From the low/medium resolution spectra observed so
far, it is already very clear that GRBs have a very different
(i.e. much stronger absorption) heavy element pattern than other high
redshift absorbers \cite{ref:savaglio,ref:savaglio2;ref:fiore}.  Among many
other GRB fundamental parameters, the Swift era will reveal whether
the strong absorption is typical of the ISM in high-$z$ star forming
galaxies, or it is a peculiarity of the GRB circumburst medium only.

\end{document}